\def\aj{AJ}

\def\mnras{MNRAS}

\def\apj{ApJ}
\def\apjl{ApJL}

\def\parn{\par\noindent}

\catcode`\@=11
\def\gsim{\ifmmode{\mathrel{\mathpalette\@versim>}}
    \else{$\mathrel{\mathpalette\@versim>}$}\fi}
\def\lsim{\ifmmode{\mathrel{\mathpalette\@versim<}}
    \else{$\mathrel{\mathpalette\@versim<}$}\fi}
\def\@versim#1#2{\lower 2.9truept \vbox{\baselineskip 0pt \lineskip 
    0.5truept \ialign{$\m@th#1\hfil##\hfil$\crcr#2\crcr\sim\crcr}}}
\catcode`\@=12
\font\boldsym=cmmib10\font\boldsym=cmmib10

\def\Omegab{{\hbox{\boldsym\char'012}}}

\def\csib{{\hbox{\boldsym\char'030}}}

\def\omegab{{\hbox{\boldsym\char'041}}}

\def\euv{{\bf e}_1}
\def\edv{{\bf e}_2}
\def\etv{{\bf e}_3}

\def\xv{{\bf x}}

\def\xiv{x_i}
\def\xjv{x_j}
\def\xkv{x_k}
\def\dxv{\dot\xv}

\def\cv{\csib}

\def\dcv{\dot\cv}

\def\Rcm{{\bf R}_{\rm CM}}
\def\Vcm{{\bf V}_{\rm CM}}
\def\Acm{{\bf A}_{\rm CM}}
\def\Om{{\cal O}}
\def\Omt{\Om^{\rm T}}
\def\Omph{\Om_1(\varphi)}
\def\Omth{\Om_2(\vartheta)}
\def\Omps{\Om_3(\psi)}

\def\OmthT{\Omt_2(\vartheta)}
\def\OmpsT{\Omt_3(\psi)}
\def\Cm{{\cal C}}
\def\Cmt{\Cm^{\rm T}}
\def\cth{\cos\vartheta}
\def\sth{\sin\vartheta}
\def\cph{\cos\varphi}
\def\sph{\sin\varphi}
\def\cps{\cos\psi}
\def\sps{\sin\psi}
\def\Ome{\Omegab}
\def\dOme{\dot\Ome}
\def\ome{\omegab}
\def\dome{\dot\ome}
\def\omeph{\ome_{\varphi}}
\def\ometh{\ome_{\vartheta}}
\def\omeps{\ome_{\psi}}
\def\dphi{\dot\varphi}
\def\dthe{\dot\vartheta}
\def\dpsi{\dot\psi}
\def\ddphi{\ddot\varphi}
\def\ddthe{\ddot\vartheta}
\def\ddpsi{\ddot\psi}
\def\dtx{d^3\xv}
\def\dtc{d^3\cv}
\def\dij{\delta_{ij}}
\def\eps{\epsilon}
\def\Tr{{\rm Tr}}
\def\lag{{\cal L}}
\def\Eext{{\bf g}}
\def\tm{{\cal T}}
\def\DT{\Delta T}
\def\amom{{\bf L}}

\def\fmom{{\bf N}}
\def\fmomi{N_i}
\def\rs{\rho}
\def\rsm{\bar\rs}
\def\rsrb{\rs_{\rm RB}}
\def\DI{\Delta I}
\def\rsg{\rs_{\rm g}}
\def\mrsg{\bar\rsg}
\def\Mg{M_{\rm g}}
\def\hg{h_{\rm g}}
\def\rsc{\rs_{\rm c}}

\def\rc{r_{\rm c}}
\def\Torb{P_{\rm orb}}
\def\Tdyn{P_{\rm dyn}}
\def\Tosc{P_{\rm osc}}
\def\uas{u_{\rm a}}
\def\uc{u_{\rm c}}
\def\epsdu{\epsilon_{21}}
\def\epstu{\epsilon_{31}}
\def\msun{M_{\odot}}

\newcommand{\eq}[1]{eq.\ (\ref{#1})}

\documentstyle{celmec} 

\def\refj{\parn\hangindent=1truecm}

\runningtitle{RIGID BODY IN A TIDAL FIELD}
\runningauthor{L. CIOTTI and G. GIAMPIERI}

\begin{opening}
\title{MOTION OF A RIGID BODY IN A TIDAL FIELD:}
\subtitle{An Application to Elliptical Galaxies in Clusters}

\author{L. CIOTTI}
\institute{Osservatorio Astronomico di Bologna, via Zamboni 33, 
I-40126 Bologna, Italy}
\author{G. GIAMPIERI}
\institute{Queen Mary and Westfield College, Astronomy Unit, 
London E1 4NS, UK}

\date{January 26, 1998}

\end{opening}

\begin{document}



\begin{abstract}

We investigate the motion, near the equilibrium configurations, of an
initially spinless rigid body subject to an external tidal field. Two
cases are considered: when the center of mass of the body is at rest
at the equilibrium point of the field generated by a generic mass
distribution, and when it is placed on a circular orbit subject to a
spherically symmetric potential. A complete analysis of the
equilibrium configurations is carried out for both cases.  First, we
derive the conditions for the stable equilibria, and then we analyze
the frequencies of oscillations around the equilibrium positions. In
view of these results, we consider the problem of alignment of
galaxies in clusters. After estimating the period of the oscillations
induced on the galaxies by the tidal field of the cluster, we discuss
the possible effect of resonances between stellar orbits inside the
galaxy and the oscillations of the galaxy as a whole; this may be a
mechanism responsible for producing an intracluster stellar
population.


\end{abstract}

\section{Introduction}

Informations about the structure and evolution of elliptical galaxies
(Es) are commonly inferred from their morphology. Roughly speaking,
this is justified by the fact that their internal two-body relaxation
time is longer than the Hubble time, so that the shape and internal
dynamics of such systems probably reflect the conditions present
immediately after the end of their formation process (Binney \&
Tremaine 1987, hereafter BT87). However, a problem with this scenario
is that Es in clusters are not isolated objects, being subject to the
effects of their environment. Since these effects are acting on time
scales shorter than the Hubble time, it is of the utmost importance to
be able to isolate them when using the present morphology to infer the
conditions at the galaxy formation epoch. In particular, the cluster's
tidal field (TF) acts continuously, and affects in general any galaxy
in the cluster. Thus, although the TF is not as strong as other
interactions (e.g., galaxy-galaxy encounters), its effects can become
important on the long term.

A first possibile effect of the TF is a modification of the shape of
galaxies belonging to the cluster itself. Ciotti \& Dutta (1994,
hereafter CD94), using N-body numerical simulations, have shown that
this is not the case, since Es behave like rigid bodies in the cluster
TF, at variance with disk galaxies (Valluri 1993). Such a conclusion
seems to be confirmed by observational results, because no systematic
differences are found between cluster and field Es (for a review see
CD94 and references therein).

A second possible effect on galaxies is their alignment with the
cluster major axis, or with the radial direction in the case of a
nearly spherical cluster. To this respect, Hawley \& Peebles (1975)
reported {\it ``a possible indication that the galaxies are
preferentially aligned along the radius vector to the center of the
{\rm [Coma]} cluster''}; this result was confirmed by Thompson
(1976). MacGillivray \& Dodd (1979a,b) observed an analogous alignment
for Es in the cluster centered on NGC439; in this case they found also
a number of galaxies with their major axis {\it perpendicular} to the
radial direction of the cluster. The same authors, observing another
spherical cluster, found a preferential alignment of Es' major axes
with the cluster's radial direction. Adams, Strom \& Strom (1980)
found a general trend for Es to be aligned with the cluster's major
axis in 7 very elongated clusters. They also reported {\it ``a small
but significant number of Es with their major axes perpendicular to
the cluster's major axis''}.  Fong, Stevenson \& Shanks (1990) found a
moderately significant radial alignment of Es, together with a less
significant tendency of Es to be oriented {\it perpendicularly} to
this direction. The best evidence for alignment is for the Brightest
Cluster Ellipticals: Trevese, Cirimele \& Flin (1992) described a {\it
``strong alignment of the brightest galaxy major axis with the long
axis of the parent cluster''}.

A third effect strictly related to the previous one is the possible
{\it resonant} interaction between stellar orbits inside the galaxies
and the cluster TF: if this is the case then we can speculate on the
possibility of {\it collisionless evaporation} of stars from galaxies
in clusters.

The N-body simulations of prolate Es in radial and circular orbits
inside clusters show that an alignemnt, due to the cluster TF, can in
fact be expected (CD94). In this paper we establish on more firm
grounds the analytical discussion of the results of CD94 concerning
the alignment. In particular, we discuss the equilibrium positions of
a triaxial galaxy at the center of a triaxial cluster, and of a
triaxial galaxy in circular orbit inside a spherical cluster,
analyzing also the stability of these equilibrium configurations. For
small displacements near the stable positions we obtain the
oscillation frequencies, which are then compared with the internal
orbital periods of the stars. Throughout our discussion (following the
N-body results of CD94), a galaxy is described as a rigid body (RB)
placed in an external gravitational field.  By linearizing the
external field, the problem is reduced to that of the integration of
the equations of motion for a RB in a tensorial force field,
corresponding to the TF. Although this problem is not integrable, one
can still determine the equilibrium positions and their stability,
along with the proper frequencies of the system.

The paper is organized as follows. In Section 2 we assume that the
center of mass of the RB is at rest inside an unspecified external
density distribution. In Section 3 we study the more complicate case
of a RB on a circular orbit inside an external, spherically symmetric
potential. In both cases, we first determine the equilibrium
configurations, and then we study their stability.  For the stable
positions the frequencies are determined analitically. Finally, in
Section 4 we discuss the astronomical application of the previous
analysis to the problem of the alignment of galaxies inside galaxy
clusters, and of galaxies on circular orbits within spherical
clusters. In addition, we compare the mean stellar orbital times
inside galaxies with the frequencies induced by the TF, and we
consider the possibility of a collisionless evaporation induced by
resonances.

\section{Body at rest at the equilibrium point of an
         external potential}

We start discussing the simple case of an initially spinless RB with
its center of mass at rest in an equilibrium point of an external
potential. We introduce an inertial reference system $C$, the
time-independent density distribution $\rs(\xv)$ producing the
external field $\Eext=-\nabla\phi(\xv)$.

Without loss of generality we can assume that the equilibrium point of
the potential is at $\xv=0$. We also assume that the characteristic
dimensions of the RB are much smaller than the typical length scale
over which the external potential changes. Thus, near the equilibrium
position
\begin{equation}
\phi(\xv)=\phi(0)-{1\over 2} \langle\tm\xv ,\xv\rangle +O(||\xv||^3),
\label{eq:field}
\end{equation}
where the brackets indicates the usual inner product in $\Re^3$ and
\begin{equation}
\tm_{ij}\equiv-\left. 
{\partial^2\phi(\xv)\over\partial\xiv\partial\xjv}\right|_{\xv=0}
\label{eq:tidal}
\end{equation}
is the TF tensor.  Note the minus sign in eqs.\ (\ref{eq:field}) -
(\ref{eq:tidal}), due to the fact that the TF is defined as the
truncation to the first order of the series expansion of $\Eext$. We
call $\rsrb(\xv;t)$ the density distribution of the RB of total mass
$M$, and we assume that the RB center of mass $\Rcm$ is initially at
rest at $\xv=0$.  The acceleration of the center of mass is given by
\begin{equation}
M\Acm=\int\Eext\rsrb\dtx \simeq M\Eext (\Rcm) 
      +\tm (\Rcm)\int (\xv-\Rcm)\rsrb \dtx.
\label{eq:Acm}
\end{equation}
It is evident that the last integral vanishes, i.e., the motion of
$\Rcm$ is not perturbed in the TF approximation; it is only in higher
order terms of the force expansion that the coupling appears. As a
result of the initial conditions imposed, only rotational motions
around $\xv=0$ are possible.  As usual we indicate with
\begin{equation}
\Im_{ij}(t)=\int(||\xv||^2\dij-\xiv\xjv)\rsrb(\xv;t)\dtx, 
\label{eq:inertia}
\end{equation}
the generic component of the RB inertia tensor $\Im$, as seen from $C$.

\subsection{Equilibrium Configurations}

In order to find the equilibrium configurations of the RB in the TF we
make use of the second cardinal equation of Dynamics, and we search
for the vanishing of the torque due to the TF. The tidal torque
$\fmom$ is obtained by the series expansion of its integral
definition:
\begin{equation}
\fmomi=-\eps_{ijk}\int\xjv{\partial\phi(\xv)\over\partial\xkv}\rsrb(\xv;t)\dtx,
\label{eq:torque}
\end{equation}
where $\eps_{ijk}$ is the Levi-Civita tensor (here and in the following, the 
summation over repeated indices is assumed).
Up to now no assumptions have been made about the orientation of the
TF with respect to $C$. Without loss of generality we can assume that
$\tm$ is in its diagonalized form, so that, using \eq{eq:field} and
the antisymmetry of $\eps_{ijk}$, \eq{eq:torque} can be rewritten as
\begin{equation}
\fmomi=\sum_{j,k=1}^3{\eps_{ijk}\over 2}\Im_{jk}\DT_{jk},
\label{eq:torque3}
\end{equation}
where $\DT$ is the {\it antisymmetric} tensor of components
$\DT_{ij}\equiv T_i-T_j$, and $T_i$ are the three principal values of
the TF.

\mbox{}From \eq{eq:torque3} it is evident that the torque along a
particular axis is affected only by the {\it difference} between the
TF components along the two orthogonal directions. It is also obvious
that if the external gravitational field is spherically symmetric
around $\xv=0$, or if the RB itself is spherically symmetric, then no
net momentum acts on the RB, i.e., all its positions are of
equilibrium.

We use \eq{eq:torque3} to find the equilibrium configurations of the
RB.  Let us define an orthogonal reference system $C'$ with its axes
directed along the principal axes of inertia of the RB, and introduce
the orthogonal transformation matrix $\Om (t)$ between $C'$ and
$C$. In $C'$ the RB is at rest, and its inertia tensor is diagonal,
with the three principal components $(I_1,I_2,I_3)$. We consider only
a non-degenerate situation (i.e., $I_1\ne I_2\ne I_3$ and $T_1\ne
T_2\ne T_3$). If $\Om$ acts on the vectors $\cv$ of $C'$ as
$\xv=\Om\cv$, then the following relation holds:

\begin{equation}
\Im_{jk}(t)=\sum_{\mu=1}^3\Om_{j\mu}(t)\Om_{k\mu}(t)I_{\mu}.  
\label{eq:inertia2}
\end{equation}
\mbox{}From \eq{eq:torque3}, assuming $\DT_{jk}\ne 0$, we conclude
that the RB is in equilibrium if and only if $\Im_{jk}=0$ for $j\neq
k$, i.e., if and only if $\Im$ is diagonal.  This is true if and only
if $C$ and $C'$ are coincident (apart from a renaming of the RB
principal axes of inertia).  Thus, {\it all the equilibrium positions
correspond to the RB inertia axes oriented along the TF principal
directions.}

\subsection{Stability}

We now consider the Lagrangian $\lag=T-U$ associated to our
problem. Expanding the external potential as in \eq{eq:field}, from
\eq{eq:tidal} and \eq{eq:inertia2} one obtains
\begin{equation}
U=\int\phi(\xv)\rsrb(\xv;t)\dtx\simeq
M\phi(0)+\pi G\rs (0)\Tr[\Im]+{1\over 2}\Om_{i\mu}^2(t) T_iI_{\mu},
\label{eq:potential}
\end{equation}
where $\Tr[\Im]$ is the trace of the RB inertia tensor, and the
property of diagonality of $\tm$ in $C$ has been exploited. The first
two terms in the r.h.s.\ of \eq{eq:potential} are constant and from
now on they will be dropped.
Before we proceed, it is necessary to write explicitely the
transformation matrix $\Om$ between $C$ and $C'$. Instead of using the
usual representation of $\Om$ in terms of the Euler angles (i.e., the
choice of the 3-1-3 rotations)\footnote{This choice complicates the
treatment due to the indeterminacy of the angles $\varphi$ and $\psi$
for a null inclination ($\vartheta=0$).} we adopt the
counter-clockwise rotations 1-2-3, namely
\begin{equation}
\Om=\Omph\Omth\Omps,
\label{eq:rotation}
\end{equation}
with
\begin{equation}
\Omph=\left(\matrix{
                        1    &      0     &       0     \cr
                        0    &    \cph    &    -\sph    \cr
                        0    &    \sph    &     \cph    \cr
                     }\right),
\end{equation}
\begin{equation}
\Omth=\left(\matrix{
                      \cth   &      0     &     \sth    \cr
                        0    &      1     &       0     \cr
                     -\sth   &      0     &     \cth    \cr
                     }\right),
\end{equation}
\begin{equation}
\Omps=\left(\matrix{
                      \cps   &   -\sps    &       0    \cr
                      \sps   &    \cps    &       0    \cr
                        0    &      0     &       1    \cr
                     }\right).
\end{equation}

The angular velocities associated with the rotations around the three
inertia axes are $\omeph^{\rm T}=(\dphi,0,0)$, etc, where the suffix
`T' means `transpose'. By vector composition, the angular velocity of
the RB, as seen from $C'$, is:
\begin{equation}
\ome =\omeps+\OmpsT [\ometh+\OmthT\omeph].
\label{eq:omega}
\end{equation}

The kinetic energy of a RB with its center of mass kept fixed is given
in $C'$ by
\begin{eqnarray}
2T & = & I_1(\dthe\sps+\dphi\cth\cps)^2+
         I_2(\dthe\cps-\dphi\cth\sps)^2+
         I_3(\dpsi+\dphi\sth)^2. \nonumber
\end{eqnarray}
Similarly, the potential energy  $U$ can be expressed as:
\begin{eqnarray}
2U    & = & (\cph\cps-\sph\sth\sps)^2\DT_{21}\DI_{21}+
            \sph^2\cth^2\DT_{21}\DI_{31}+\nonumber\\
      &   & (\sph\cps+\cph\sth\sps)^2\DT_{31}\DI_{21}+
            \cph^2\cth^2\DT_{31}\DI_{31}.\nonumber\\
\end{eqnarray}
where we have introduced the antisymmetric tensor $\Delta_{ij}=
I_i-I_j$, and neglected an additive constant.

As proved in \S 2.1, apart from a renaming of the inertia axes, the
equilibrium positions are given by
$(\varphi,\vartheta,\psi)_{eq}=(0,0,0)$. The linearized kinetic and
potential energy near this point are simply
\begin{eqnarray}
T &\simeq& {1\over 2}I_1\dphi^2+{1\over 2}I_2\dthe^2+{1\over 2}I_3\dpsi^2,\\
U &\simeq& -{1\over 2}\DT_{32}\DI_{32}\varphi^2
     -{1\over 2}\DT_{31}\DI_{31}\vartheta^2
     -{1\over 2}\DT_{21}\DI_{21}\psi^2.
     \label{eq:potrid}
\end{eqnarray}
Note how both the quadratic forms are already diagonal: this means
that the chosen coordinates (angles) are also the normal coordinates
for the problem. The differential equations of the motion are readily
obtained:
\begin{eqnarray}
\ddphi & = &    {\DT_{32}\DI_{32}\over I_1}\varphi     
\label{eq:motiona}\\
\ddthe & = &    {\DT_{31}\DI_{31}\over I_2}\vartheta   
\label{eq:motionb}\\
\ddpsi & = &    {\DT_{21}\DI_{21}\over I_3}\psi ,  
\label{eq:motionc}     
\end{eqnarray}
and their solution is trivial.  Note that in the determination of
stability positions and relative frequencies, only the {\it ratios} of
the inertia moments matter.  Accordingly, in order to simplify our
successive discussion in \S 3, we define
\begin{equation}
u\equiv{I_1\over I_3};\quad v\equiv{I_2\over I_3}.
\label{eq:uv}
\end{equation}
Without loss of generality, let us assume that $T_1\geq T_2\geq T_3$,
i.e., that $\DT_{32}$, $\DT_{31}$, and $\DT_{21}$ are all less or
equal to zero.  Thus, from eqs.\
(\ref{eq:motiona})-(\ref{eq:motionc}), stable motions are possible if
and only if $1>v>u$.  In conclusion, a {\it stable} equilibrium
configuration (in the non-degenerate case) is achieved when:
\begin{enumerate}
\item The principal axes of the the TF are superimposed with those of the RB;
\item If $T_1\geq T_2\geq T_3$ then $I_1\leq I_2\leq I_3$. 
\end{enumerate}
For what concern the behavior of the frequencies, it is obvious that
their values increase if we increase the differences between the TF
eigenvalues $T_i$ and/or the values of the inertia parameters $u,v$.
In \S 4 the physical meaning of points 1 and 2 above will be discussed
for our astronomical application, and an explicit calculation of some
typical frequencies will be derived for galaxy and clusters models
matching observational constraints.

\section{Body on a circular orbit in a spherical potential}

In this section we discuss the case of a triaxial RB on a circular
orbit inside a spherically symmetric potential. Note that the
assumption that the RB's orbit remains circular is justified by the
fact that the external field, truncated at the TF approximation, does
not affect the motion of $\Rcm$ (see \S2).

Without loss of generality, we can assume that the orbit of $\Rcm$ is
in the $(x,y)$ plane of the inertial system $C_0$, the external
density is given by $\rs=\rs(r)$ and its associated mass inside the
radius $r$ is $M(r)$. The natural frame for our problem (Fig.\ 1) is
the {\it non-inertial} reference system $C=(O;\euv,\edv,\etv)$ with
its origin placed at $\Rcm$, $\euv=\Rcm/||\Rcm||$, and $\etv$ parallel
to the $z$-axis of $C_0$. As a consequence $C$ is rotating with the
same angular velocity $\Ome$ of $\Rcm$:
\begin{equation} 
\Ome^{\rm T}=(0,0,\Omega), \qquad 
\Omega^2=\left.{GM(r)\over r^3}\right|_{r=R_{\rm CM}}.  
\end{equation} 
We indicate with $\cv$ and $\dcv$ respectively the generic position
and velocity vector in $C$, related to $\xv$ and $\dxv$ by the usual
transformation rules involving the orthogonal transformation matrix
$\Cm$ between $C_0$ and $C$ (see, e.g., Arnold 1978).
\begin{figure} 
\vspace{5cm} 
\caption{The three reference systems used in the discussion of the RB 
         in circular motion inside a spherical potential.}  
\label{refsys} 
\end{figure}
\subsection{Equilibrium configurations}

As in the previous case we start discussing the RB equilibrium
positions, by means of the second cardinal equation of Dynamics. The
relations between the angular momenta and torques in the two systems
are given by
\begin{equation}
\amom_0 = M\Rcm\wedge\Vcm+\Cm (\amom+\Im\Ome),
\end{equation}
and
\begin{equation}
\fmom_0=M\Rcm\wedge\Acm+\Cm\fmom,
\end{equation}
where $\Im$ is the (time-dependent) RB inertia tensor in $C$, and 
\begin{equation}
\amom=\int\cv\wedge\dcv\rsrb\dtc ,
\quad\fmom=\int\cv\wedge (\Cmt\Eext)\rsrb\dtc.
\end{equation}
\mbox{}From $\dot\amom_0=\fmom_0$ one obtains, upon multiplication by $\Cmt$:
\begin{equation}
\dot\amom=\fmom-\dot\Im\Ome-
\Im\dot\Ome-\Ome\wedge\amom- \Ome\wedge\Im\Ome.
\label{eq:equilibrium}
\end{equation}

The equilibrium positions are determined by the request that
$\dot\amom=\amom=\dot\Im=0$, along with the condition (verified by
hypothesis) that $\dOme=0$. Eq.\ (\ref{eq:equilibrium}) then gives
\begin{equation}
\fmom-\Ome\wedge\Im\Ome=0.
\label{eq:equilibrium2}
\end{equation}

The first term to be discussed is the tidal torque $\fmom$. A
treatment analogous to that described in \S 2 gives again
\eq{eq:torque3}, where now (CD94)
\begin{equation}
\tm=-{4\pi G\rsm\over 3}
     \left(\matrix{
                    3q-2     &    0   &  0     \cr
                     0       &    1   &  0     \cr
                     0       &    0   &  1     \cr
                     }\right),
\label{eq:cd}
\end{equation}
having defined
\begin{equation}
q(r)\equiv {\rho(r)\over\rsm(r)}\leq 
           1\footnote{If $d\rho/dr\leq 0$ then the inequality $q\le1$ is 
                      trivial to prove. If $\rho(r) \propto r^{-\alpha} 
                      (\alpha<3)$, then $q=1-\alpha/3$. Outside a spherical 
                      body $q=0$.},
\label{eq:q}
\end{equation}
where
\begin{equation}
\rsm(r)\equiv {3M(r)\over 4\pi r^3}={3\Omega^2(r)\over 4\pi G}.
\end{equation}

Note how in this case the TF is degenerate, i.e., two of three TF
principal components are equal: $T_2=T_3$ and so $\DT_{23}=0$; also,
$\DT_{12} = \DT_{13}$.

We can now proceed with the discussion of  \eq{eq:equilibrium2}, which gives
\begin{eqnarray}
\Im_{23}\Omega^2            & = & 0,
\label{eq:eq3a}\\
\Im_{13}(\DT_{13}+\Omega^2) & = & 0,
\label{eq:eq3b}\\
\Im_{12}\DT_{12}            & = & 0.
\label{eq:eq3c}
\end{eqnarray}
Since from eqs.\ (\ref{eq:cd}) and (\ref{eq:q}) we get
\begin{equation} 
\DT_{13}=\DT_{12}=3(1-q)\Omega^2\geq 0,
\end{equation} 
eqs.\ (\ref{eq:eq3a}) - (\ref{eq:eq3c}) give the same conditions as in
the previous section, namely the equilibrium is achieved when the
off-diagonal components of $\Im$ vanish in $C$, which in this case
means that one inertia axis is directed radially, and another is
directed along the angular velocity direction. Thus, $\Ome$ removes
the intrinsic degeneracy of the TF.

\subsection{Stability}

We will now linearize $\lag$ around the equilibrium points. Let $C'$
be the reference system whose axes are the principal axes of the RB,
and $\Om$ the transformation matrix between $C$ and $C'$ [see Fig.\ 1
and \eq{eq:rotation}]. The final expression for the kinetic energy is:
\begin{equation}
T={M V_{\rm CM}^2\over 2}+
{\langle\Ome'+\ome,I(\Ome'+\ome)\rangle\over 2},
\label{eq:kinetic}
\end{equation}
where $\ome$, given by \eq{eq:omega}, is the angular velocity of the
RB as seen from $C'$, and $\Ome'=\Omt\Ome$.

Obviously, the first term in the r.h.s.\ of the previous equation is
constant, and so can be discarded. Equation (\ref{eq:kinetic})
contains a term linear in the generalized velocities, a well known
consequence of the time dependence of the new coordinate system. With
a careful analysis, however, it is easy to prove that this linear
component is identically null for a constant $\Ome$, so that it does
not affect the motion. The potential energy remains unaffected by the
coordinate change, and its expression is again given by eq.\
(\ref{eq:potrid}). The differential equations associated with the
quadratic part of $\lag$ are:
\begin{eqnarray} 
\ddphi & = &    -{\Omega^2\DI_{32}\over I_1}\varphi+
                 {\Omega(I_1+I_2-I_3)\dthe\over I_1}, 
\label{eq:motion2a}\\
\ddthe & = &    -{(\Omega^2+\DT_{13})\DI_{31}\over I_2}\vartheta-
                 {\Omega(I_1+I_2-I_3)\dphi\over I_2}, 
\label{eq:motion2b}\\
\ddpsi & = &    {\DT_{21}\DI_{21}\over I_3}\psi ; 
\label{eq:motion2c}      
\end{eqnarray}
note that for $\Omega=0$ we reobtain 
eqs.\ (\ref{eq:motiona})-(\ref{eq:motionc}) (with $\DT_{23}=0$).

We now analyze the stability of the system described by eqs.\
(\ref{eq:motion2a})-(\ref{eq:motion2c})\footnote {These linearized
equations can also be obtained by expressing \eq{eq:equilibrium} in
$C'$:
$$I\dome=\fmom'-(\Ome'+\ome)\wedge I(\Ome'+\ome)-I(\Ome'\wedge\ome).$$ 
Note the identical formal aspect of this system with that discussed in
BT87, p.138 for the stability of Lagrangian points.}.  The first
result follows directly from \eq{eq:motion2c}, from which, by virtue
of the assumption $\DT_{21}\leq 0$, we get the stability condition
$I_2\geq I_1$.

Equations (\ref{eq:motion2a})-(\ref{eq:motion2b}), after some standard
manipulations, are separated into two linear fourth-order equations
with constant coefficients, that are solved setting
\begin{equation}
\varphi (t)\equiv X\exp(\lambda t),\quad 
\vartheta (t)\equiv Y\exp(\lambda t).
\end{equation}
The resulting characteristic equation for $\lambda$ is
biquadratic\footnote{The separation introduces two additional
integration constants for each variable. These are determined by the
requirement that eqs. (\ref{eq:motion2a})-(\ref{eq:motion2b}) are
satisfied at each instant of time.}:
\begin{equation}
\lambda^4 +\alpha\lambda^2+\beta=0,
\label{eq:characteristic}
\end{equation}
where 
\begin{eqnarray}
\alpha &\equiv& \Omega^2\left[ 
{(u+v-1)^2\over uv}+{1-v\over u}+{(4-3q)(1-u)\over v}\right]\\
\beta &\equiv& \Omega^4{(4-3q)(1-u)(1-v)\over uv},
\end{eqnarray}
and $u$ and $v$ are given in eq. (\ref{eq:uv}). Any solution of 
\eq{eq:characteristic} is of the form
\begin{equation}
\lambda=\Omega\tilde\lambda(u,v;q),
\end{equation}
where $\tilde\lambda$ is a dimensionless function, and the dependence
on $r$ is contained in $\Omega$ and $q$. This means that in the
determination of stable equilibria only $(u,v;q)$ matter, i.e., the
parameter space is three--dimensional. Moreover, for fixed $(u,v;q)$
the frequencies are proportional to $\Omega$, the orbital velocity of
$\Rcm$ around the center of the external density distribution.

Equation (\ref{eq:characteristic}) has four solutions: if $\lambda$ is
a root also $-\lambda$ is a root, and so the general solution will
contain an exponential increasing term if $\Re{\rm e}(\lambda)\neq 0$,
making it unstable.  The only possibility for a stable equilibrium is
then for the four solutions to be purely imaginary, and this is
possible if and only if the two solutions of the associated quadratic
equation are real and negative. The reality requires that
\begin{equation}
\Delta\equiv \alpha^2-4\beta >0,
\label{eq:req1}
\end{equation}
and for the negativity it is required that
\begin{equation}
\alpha>0;\quad \beta>0.
\label{eq:req2}
\end{equation}
Note that the conditions (\ref{eq:req1})-(\ref{eq:req2}) are formally
equivalent to the single condition $\alpha>2\sqrt{\beta}$.  The
positivity of $\beta$ is obtained in two different cases:
\par\smallskip
      (a) $u<v<1$, i.e., $I_3>I_2>I_1$.
\par\smallskip
      (b) $1<u<v$, i.e., $I_2>I_1>I_3$.
\par\smallskip\noindent
where the stability of $\psi$ motion ($I_2>I_1$) has been already
assumed.

Case (a) is of easy discussion: in fact, $\alpha $ is a sum of
positive quantities, and $\Delta$ can be rewritten as a sum of
positive quantities as well, independently of the value of $q(r)$.
Thus, {\it the motion around the equilibrium position is stable for
$I_3>I_2>I_1$, for any given density profile of the external density
distribution}. This corresponds to region (a) in Fig.\ 2.

Case (b) is far more interesting. Its discussion is algebrically
cumbersome, but not intrinsecally difficult. It can be proved that
when $1<u<v$ the request $\Delta >0$ implies $\alpha>0$, so that the
only condition to be studied is $\Delta >0$. The stability region in
the $(u,v)$ plane is plotted in Fig.\ 2 for a specific value of $q$.
\begin{figure}
\vspace{5cm}
\caption{The $(u,v)$ plane with the various critical curves, for $q=0.9$. 
         In region $a$ the major axis of the RB is directed toward the center 
         of the external potential, the intermediate axis is directed 
         tangentially to the orbit and the minor axis is perpendicular to the 
         orbital plane (Fig. 1). In region $b = b_1 \cup b_2$ the major axis 
         of the RB is perpendicular to the orbital plane, the medium axis is 
         directed toward the center, and the short axis is parallel to the 
         orbital velocity.}
\label{stabreg}
\end{figure}
Note that the stability region relative to case (b) is made of two
qualitatively different parts, separated by a vertical asymptote,
located at:
\begin{equation}
\uas (q)={1\over 2}+{1\over 2}\sqrt{{4-3q\over 3(1-q)}}.
\end{equation}
In region $b_1$ ($1<u<\uas$, see Fig.\ 2) there is no upper bound on
$I_2/I_3$, and $q\to 1 \Rightarrow \uas\to\infty$, i.e., region $b_1$
expands to all $v>u>1$.  For any finite value of $q$, there exist
another stability region ($b_2$, see Fig.\ 2) to the right of $\uas$,
whose lower bound is again represented by $v=u$. The upper bound on
$I_2/I_3$ (determined by the condition $\Delta =0$) is given by
\begin{equation}
v_+=(u-1){1+9(1-q)u-6(1-q)u^2+6(1-q)u\sqrt{u(u-1)(4-3q)}\over
         12(1-q)u^2-12(1-q)u-1},
\end{equation}
and the point of intersection between $v_+(u)$ and $v=u$ is given by 
\begin{equation}
\uc (q)={1\over 2}+{1\over 2}\sqrt{{-9q^2+6q+11+8\sqrt{4-3q}\over 
                                   3(1-q)(5+3q)}}.
\end{equation} 
The functions $\uas (q) $ and $\uc (q) $ are plotted in Fig.\ 3. A
very important case is obtained for $q=0$, i.e., in the case of a RB
orbiting {\it outside} a spherical body: in this case the region of
equilibrium is very small, since $\uas (0)=1/2+ 1/\sqrt{3}\simeq 1.08$
and $\uc (0)=1/2+3\sqrt{5}/10 \simeq 1.17$.
\begin{figure}
\vspace{5cm}
\caption{The behaviour of $\uas$ and $\uc$ (see Fig. 2) as a function of $q$.}
\label{uasuc}
\end{figure}

\subsection{Frequencies}

Having found the criteria for stable equilibrium, we now focus our
attention on the frequencies of the motions around the
($\varphi,\vartheta,\psi$) axes. The frequency for the $\psi$ motion
is obtained directly from eq. (\ref{eq:motionc}) as

\begin{equation}
{2\pi\over P_{\psi}} = 
\Omega\sqrt{k (v-u)};\quad\quad k\equiv 3(1-q). 
\end{equation}
Its value increases by increasing $\Omega$ and/or decreasing $q$,
i.e., by making the external density more concentrated.

The motion around $\varphi$ and $\vartheta$ results from the superposition
of two independent oscillations, with frequencies
\begin{equation}
{2\pi\over P^{\pm}_{\varphi}} =
{2\pi\over P^{\pm}_{\vartheta}} = 
\Omega\sqrt{{1+(\sigma + k) \sigma \pm \sqrt{ 1 +
\sigma^2 (\tau +k)^2-2\sigma (\tau +2k\tau  -  k)} \over 2}},
\end{equation}
where
\begin{eqnarray*}
\sigma \equiv \frac{1-u}{v}; \quad\quad \tau\equiv \frac{1-v}{u}.
\end{eqnarray*}

We stress again the presence of the proportionality constant $\Omega$
for all frequencies: assuming that the dimensionless factor is of the
order of unity, the oscillatory motion around the stable positions is
characterized by periods of the order of the RB orbital time.

\section{An Astrophysical Application: Galaxies in Clusters}

We now apply the previous results to the cases of a triaxial galaxy at
the centre of a triaxial cluster, and of a triaxial galaxy in circular
orbit inside a spherical cluster. We start presenting the galaxy model
used.

\subsection{The galaxy model}

We assume that the galaxy density is stratified on homeoidal surfaces
labeled by
\begin{equation}
m^2=\sum_{i=1}^3{x_i^2\over\alpha _i^2},
\quad\alpha_1\geq\alpha_2\geq\alpha_3,
\label{eq:homo}
\end{equation}
i.e., $\rsg=\rsg(m)$. In the reference system implicitely assumed in
\eq{eq:homo}, all the non-diagonal terms of $\Im$ vanish, and the
diagonal terms are given by:
\begin{equation}
I_i =  {4\pi\over3}\alpha_1\alpha_2\alpha_3(\alpha_j^2 +\alpha_k^2)\hg;\quad
i\neq j\neq k,
\label{eq:I_i}
\end{equation}
where the form factor is 
\begin{equation}
\hg\equiv\int_0^{\infty}\rsg(m)m^4dm.
\end{equation}

Note that the inertia moments satisfy $I_1\leq I_2\leq I_3$. Note also
that in all our results in \S 2 and \S 3 only the quantities
$u=I_1/I_2$ and $v=I_2/I_3$ appear, and in particular we conclude that
{\it for a triaxial homeoidal body the equilibrium positions and the
relative frequencies do not depend on the quantity $\hg$, i.e., they
do not depend on the particular density distribution under
consideration}.

However, a particular density distribution for the galaxy model has to
be assumed in order to estimate the characteristic stellar orbital
times, and so we adopt an ellipsoidal generalization of the widely
used $\gamma$-models (Denhen 1993, Tremaine et al. 1994):
\begin{equation}
\rsg(m)={\Mg\over\alpha_1\alpha_2\alpha_3}{3-\gamma\over 4\pi}
        {1\over m^{\gamma}(1+m)^{4-\gamma}},
\label{eq:densitylaw} 
\end{equation} 
where $0\leq\gamma <3$. The density in the outer parts goes like
$m^{-4}$ independently of $\gamma$; for $\gamma=0$ the models present
a flat density {\it core}. It is straightforward to show that a
density stratified on constant $m$ surfaces, when projected on its
principal planes, originates a brightness distribution with elliptical
isophotes whose ellipticity is the same as that of the corresponding
section of the ellipsoid. For example, projecting along $x_3$ the
isophotal axial ratio in the $x_1-x_2$ plane is $\alpha_2/\alpha_1$,
and the corresponding ellipticity is $\epsdu
=1-\alpha_2/\alpha_1$. This property is useful when constructing
realistic galaxy models: mean observed values are $\eps=0.2-0.4$
(corresponding to E2-E4 galaxies).

For the models described by eq. (\ref{eq:densitylaw}) the mass inside
$m$ is given by:
\begin{equation}
\Mg(m)=4\pi\alpha_1\alpha_2\alpha_3\int_0^m\rsg(t)t^2dt=
\Mg{m^{3-\gamma}\over (1+m)^{3-\gamma}},
\end{equation}
the mean density inside $m$ is
\begin{equation}
\mrsg (m) ={\Mg\over\alpha_1\alpha_2\alpha_3}{3\over 4\pi}
        {1\over m^{\gamma}(1+m)^{3-\gamma}},
\label{eq:rhom}
\end{equation}
and an {\it estimate} of the order of magnitude of the characteristic orbital 
time of stars inside $m$ is finally obtained as usual as
\begin{equation}
\Torb (m)\simeq 4\Tdyn (m)\equiv\sqrt{3\pi\over G\mrsg(m)},
\label{eq:torb}
\end{equation}
(BT87, p.37)\footnote{Note that, strictly speaking,  the previous formula
applies only to the case of a point mass in radial or circular orbit
inside a homogeneous, spherically symmetric density distribution. As
pointed out by the referee, particular attention should be paid in the
case of an incompressible, self-gravitating, prolate spheroid steadily
rotating about its axis of symmetry: in fact, in this case the
equations of hydrodynamics do not admit solutions (see, e.g., Florides
\& Spyrou 1993).}.

\mbox{}From eqs.\ (\ref{eq:rhom})-(\ref{eq:torb}) one can write
\begin{equation}
\Torb =\sqrt{{\alpha_1\alpha_2\alpha_3\over G\Mg}}\,
       \widetilde\Torb\simeq 3.0\,
       \sqrt{\alpha_{1,10}^3(1-\epsdu)(1-\epstu)\over M_{\rm g,11}}
       \widetilde\Torb (m,\gamma)\times 10^8\,{\rm yr},
\end{equation}
where $M_{\rm g,11}$ is the galaxy mass normalized to $10^{11}\msun$
and $\alpha_{1,10}$ is the galaxy core major axis in 10 kpc. In Fig.\
4 we plot the dimensionless function $\widetilde\Torb$, for various
values of $\gamma$.
\begin{figure}
\vspace{5cm}
\caption{Extimated orbital times as function of $m$ for various $\gamma$. 
         The curves from top to bottom correspond, respectively, to 
         $\gamma$ equal to 0, 1, 1.5, 2, and 2.5. The black diamonds 
         correspond to the half-mass orbital time, i.e., the orbital time
         calculated at $m_{1/2}=[2^{1/(3-\gamma)}-1]^{-1}$, the homeoid 
         containing half of the galaxy mass.}
\label{short name}
\end{figure}
\subsection{Triaxial Galaxies at the Center of Triaxial Clusters}

We now focus on the TF produced by a triaxial cluster. We assume that
the cluster's density is stratified on homeoidal surfaces
$\rsc=\rsc(\ell^2)$, with
\begin{equation}
\ell^2=\sum_{i=1}^3{x_i^2\over a_i^2};\quad a_1\geq a_2 \geq a_3.
\end{equation}

The associated potential is found using the well-known formulas from
potential theory (Kellog 1953; Chandrasekhar 1969, Cap.3; BT87 p.61).
With the aid of the auxiliary function
\begin{equation}
\Psi(\ell^2)\equiv\int_0^{\ell^2}\rsc(t)dt,
\end{equation}
the potential can be written as
\begin{equation}
\phi(\xv)=-G\pi a_1a_2a_3\int_0^{\infty}
{\Psi(\infty)-\Psi[\ell^2(\tau,\xv)]\over\Delta(\tau)}d\tau,
\end{equation} 
where
\begin{equation}
\Delta(\tau)=\sqrt{(a_1^2+\tau)(a_2^2+\tau)(a_3^2+\tau)},
\end{equation}
and
\begin{equation}
\ell^2(\tau,\xv)\equiv \sum_{i=1}^3{x_i^2\over a_i^2+\tau}.
\label{eq:elle2}
\end{equation}
The fact that the center of the cluster is an equilibrium point is
trivial to prove, and with very simple algebra the following integral
expression for the principal components of the TF at the cluster
center are found:
\begin{equation}
T_i=-2\pi G\rsc(0)a_1a_2a_3\int_0^{\infty}{d\tau
                                   \over (a_i^2+\tau)\Delta(\tau)}.
\label{eq:Ti}
\end{equation}
Note that the particular form of the cluster density distribution does
not enter eq.\ (\ref{eq:Ti}), and that if $a_1\geq a_2\geq a_3$ then
$T_1\geq T_2\geq T_3$, as assumed in \S 2.2.  We can now give the
astrophysical interpretation of the results obtained in \S 2.2: the
stable equilibrium position of a triaxial galaxy at the center of a
triaxial cluster corresponds geometrically to the three principal axes
of the galaxy to be collinear -- in the same order -- with the
principal axes of the cluster density distribution. This configuration
reproduces in a natural way the observed {\it ``strong alignment of
the [brightest] galaxy major axis with the long axis of the parent
cluster''} (Trevese, Cirimele, Flin 1992), already mentioned in the
Introduction.

In order to estimate the oscillatory periods of a galaxy around its
equilibrium position, the numerical values of the TF components in a
realistic cluster are needed. \mbox{}From eq.\ (\ref{eq:Ti}) one obtains
\begin{equation} 
\DT_{ij}=-2\pi G\rsc(0)(1-\epsdu)(1-\epstu) \Delta\widetilde T_{ij}.
\label{eq:DTi} 
\end{equation}
The evaluation of the three integrals $T_1$, $T_2$, and $T_3$ involves
elliptic functions; in the degenerate case of a rotation ellipsoid
these functions can be expressed in terms of elementary functions.
The dimensionless quantities $\widetilde T_i$ are given in Appendix.
We obtain $\rsc(0)$ in \eq{eq:DTi} using the Virial Theorem for
the King density distribution (King 1972, BT87 p.228), from
which $\rsc(0)\simeq 9\sigma^2 / 4\pi G a_1^2$. The time required for
a complete oscillation of the galaxy around each of the angular
coordinates is then given by
\begin{equation} 
\Tosc=\sqrt{{2\pi\over G\rsc(0)}}\widetilde\Tosc(u,v,\epsdu,\epstu)\simeq 
7.2\,{a_{1,250}\over\sigma_{1000}} 
\widetilde\Tosc(u,v,\epsdu,\epstu)\times 10^8 \,{\rm yr}, 
\label{eq:Tosc} 
\end{equation}
where $\sigma_{1000}$ is the central (1-dimensional) cluster velocity
dispersion normalized to $1000$ km s$^{-1}$ (a number of order unity
for rich clusters), and $a_{1,250}$ is the cluster core major axis
normalized to 250 kpc.  Following Valluri (1993) and Girardi et al.\
(1995), we adopt $a_{1,250}=1$.  The explicit expressions for the
three $\widetilde\Tosc$ can be obtained from eqs.\
(\ref{eq:motiona})-(\ref{eq:motionc}). The dimensionality of the
parameter space is too large to allow a complete exploration, so we
limit ourselves to a specific case.  For example, let us assume
$a_2/a_1=0.8$, $a_3/a_1=0.6$, and $\sigma_{1000}=1$.  For the galaxy
model, we also adopt $\alpha_2/\alpha_1=0.8$ and
$\alpha_3/\alpha_1=0.6$, corresponding to an E2 and E4 galaxy when the
model is projected on the $x_1-x_2$ and $x_1-x_3$ planes,
respectively.  With this choice of parameters, $u=I_1/I_3=25/41$ and
$v=I_2/I_3=34/41$, and the three values of $\Tosc$ are
$(2.7,1.6,3.9)\times 10^9$ yrs, for motions around $\varphi,\vartheta,
\psi$, respectively. So, a comparison with the mean stellar orbital
times obtained in the previous section (see Fig.\ 4) shows that in the
outer halo of giant Es the stellar orbital times can be of the same
order of magnitude as the oscillatory periods of the galaxies
themselves. It could be of some interest a numerical investigation of
the final fate of these halo stars: in case of escape this is a {\it
collisionless} evaporation. Note that, at variance with the
collisional evaporation of globular clusters, in this case no mass
segregation is expected, since the effect discussed here is
independent of stellar mass. On the contrary, for the bulk of stars in
galaxies the orbital times are considerably shorter than $\Tosc$, and
this is the reason why in the N-body simulations of CD94 galaxies are
found to behave like rigid bodies.

\subsection{Triaxial Galaxies in Circular Orbit in Spherical Clusters}

As shown in \S 3, in this case there are two different equilibrium
configurations. The first equilibrium position, with the galaxy's
major axis directed toward the galaxy center, reproduces the main
observational results already discussed in the Introduction.  In the
second equilibrium position the galaxy major axis is {\it
perpendicular} to the orbital plane, as observed for a small number of
cases.

In order to analyze the characteristic frequencies around the
equilibrium positions in this case, a triaxial galaxy described in \S
4.1 is assumed on circular orbit inside a spherical cluster, described
by a Hernquist (1990) density distribution [\eq{eq:densitylaw} with
$\gamma=1$, $\alpha_1=\alpha_2=\alpha_3=\rc$].  This assumption about
the mass profile follows from the most recent results of cosmological
high-resolution collisionless N-body simulations (see, e.g., White
1996, and references therein).  \mbox{}From the Virial Theorem applied
to this density law (Hernquist 1990), one finds that
$18\rc\sigma^2=GM_c$, where $\sigma$ is the virial 1-D velocity
dispersion.

The radial trend of $q(r)$ is plotted in Fig.\ 5 (upper panel, solid
line): $q$ monotonically decreases moving outwards from the cluster
center\footnote{For $\gamma$-models, $q(r)=(1-\gamma/3)/(1+r/\rc)$.},
where $q=2/3$.  In the same figure, we also plot the normalized
angular velocity $\widetilde\Omega=\Omega/\sqrt{G M_{\rm c}/\rc^3}$
(dashed line).  The equation analogous to eq.\ (\ref{eq:Tosc}) is now
\begin{equation} 
\Tosc={2\pi\over\sqrt{GM_{\rm c}/\rc^3}}\;
{\widetilde\Tosc(u,v,q)\over \widetilde\Omega}\simeq 
3.6\, {r_{\rm c,250}\over\sigma_{1000}}
{\widetilde\Tosc(u,v,q)\over \widetilde\Omega}\times 10^8\, {\rm yr}, 
\label{eq:Tosck} 
\end{equation}
where, for each angle, $\widetilde\Tosc(u,v,q)$ follows from eqs.\
(58)-(59), and is plotted in Fig.\ 5 (lower panel) as a function of
$r/\rc$ for the adopted $u$ and $v$.
\begin{figure}
\vspace{5cm}
\caption{Upper panel: the radial trend of $q$ (solid line) and
         $\widetilde\Omega$  (dotted line) for the Hernquist model. The radius 
         is normalized to the cluster core radius. Lower panel: the radial
         trend of the three characteristic $\widetilde\Tosc /\widetilde\Omega$.
         The solid line corresponds to the motion around $\psi$, and the 
         dashed and dotted lines to $\Tosc^-$ and $\Tosc^+$ respectively.}
\label{spheclu}
\end{figure}
Note how $\Tosc$ steadily increases moving outward in the cluster.  As
a final result there exists a radius approximately of the order of the
cluster core where the oscillatory times of galaxies are comparable
with their mean stellar orbital times, and so it is possible that some
stars will be affected by resonances. The fate of these orbits can be
investigated by numerical integration of the equations of motion.

\section{Conclusions}

In this paper we have presented a discussion of the motion, near the
equilibrium configurations, of a rigid body subject to a tidal field.
A complete analysis of the equilibrium points when the center of mass
of the body is at rest in the field generated by a generic mass
distribution, and when it is placed on a circular orbit inside a
spherically symmetric potential, was given. The conditions for stable
equilibria and the frequencies of oscillations around such positions
are analitically derived.  As an astrophysical application of the
previous results, we have discussed the observed alignment of galaxies
in clusters, concluding the numerical investigations presented in
CD94. In particular, we have shown that the radially aligned
configurations found through N-body simulations by CD94 are indeed
equilibrium positions.  We have also shown the existence of an
equilibrium configuration not found by CD94, i.e., the case of a
galaxy in circular orbit in an external spherical potential, whose
major axis is perpendicular to the orbital plane; such configurations
are observed for a minority of galaxies in real clusters.  Moreover,
comparing the orbital times of stars inside the model galaxies with
the oscillatory times around the equilibrium configurations of the
galaxies in the cluster tidal field, we found that for realistic
parameters, valid for the majority of the galaxies, the stellar
orbital times are much shorter, and so the orbits are adiabatically
invariant. This could be an explanation of the result numerically
found by CD94 that N-body galaxies behaves in the cluster TF as rigid
bodies.

Finally, we have pointed out the possible effect of resonances between
stellar orbits in the external parts of the galaxies, and the
oscillations frequencies induced on the galaxies by the tidal field of
the parent cluster. We found that the induced frequencies on galaxies
can be of the order of the stellar orbital times in their outer halos,
and this fact suggests the existence of a possible resonant
collisionless evaporation of stars from galaxies in clusters.  This
mechanism could explain the recently observed population of
intracluster stars (see, e.g., M\'endez et al.\ 1997, and references
therein).

\section{Acknowledgements}

LC thanks James Binney and Prasenjit Saha for useful discussions, and
the warm ospitality of the Department of Theoretical Physics of Oxford
University, where this work was partially carried out. We would like
to thank the referee, Prof.\ Spyrou, for his comments. This work was
supported by the EEC contract No. CHRX-CT92-0033 and by the contract
ASI-95-RS-152.

\appendix

We start writing explicitely the expressions for the dimensionless
functions $\widetilde T_i$, defined as $T_i\equiv -2\pi
G\rsc(0)(a_2a_3/a_1^2)\widetilde T_i$ [see \eq{eq:Ti}] in the general
triaxial case, defined by $a_1>a_2>a_3$.

\mbox{}From Byrd \& Friedman (1971, hereafter BF71), defining
\begin{equation}
\theta=\arccos\left ({a_3\over a_1}\right);\quad 
       k^2={1-a_2^2/a_1^2\over 1-a_3^2/a_1^2},
\end{equation}
one obtains
\begin{equation}
\widetilde T_1={2[F(\theta,k)-E(\theta,k)]\over k^2 \sin^3(\theta)} 
\end{equation}
(BF71, 238.03-310.02),
\begin{equation}
\widetilde T_2={2[E(\theta,k)-(1-k^2)F(\theta,k)-(a_3/a_2)k^2\sin(\theta)]
               \over k^2(1-k^2)\sin^3(\theta)},
\end{equation}
(BF71, 238.04-318.05), and
\begin{equation}
\widetilde T_3={2[(a_2/a_3)\sin(\theta) -E(\theta,k)]\over 
               (1-k^2)\sin^3(\theta)},
\end{equation}
(BF71, 238.05-316.02), where $F(\theta,k)$ and $E(\theta,k)$ are the
Elliptic Integrals of the first and second kind, respectively (see,
e.g., BF71, 110.02- 100.03).

The first degenerate case is that of a prolate cluster density, i.e.,
$a_1>a_2=a_3$. In this case, the ellipsoidal eccentricity is given by
\begin{equation}
e=\sqrt{1-{a_2^2\over a_1^2}}
\end{equation}
and the two different components of the TF are given by:
\begin{equation}
\widetilde T_1={2\over a_2^2/a_1^2}
\left[{1-e^2\over 2 e^3}\ln\left({1+e\over 1-e}\right)-{1-e^2\over e^2}\right],
\end{equation}
(Gradshteyn \& Ryzhik 1965, eq. 2.248.1; hereafter GR65), and
\begin{equation}
\widetilde T_2=\widetilde T_3={1\over a_2^2/a_1^2}
\left[{1\over e^2}-{1\over 2 e^3(1-e^2)}\ln\left({1+e\over 1-e}\right)\right],
\end{equation}
(GR65, 2.248.4).
The other degenerate case is the oblate one. We assume now $a_1=a_2>a_3$, 
and so
\begin{equation}
e=\sqrt{1-{a_3^2\over a_1^2}}.
\end{equation}
The components of the TF are readily found:
\begin{equation}
\widetilde T_1=\widetilde T_2={1\over a_3/a_1}
\left[{\sqrt{1-e^2}\arcsin (e)\over e^3}-{1-e^2\over e^2}\right],
\end{equation}
(GR65 2.248.4), and
\begin{equation}
\widetilde T_3={2\over a_3/a_1}
\left[{1\over e^2}-{\sqrt{1-e^2}\arcsin (e)\over e^3}\right],
\end{equation}
(GR65 2.248.1).

\section{References}

\refj{Arnold V.I.:1978 
     {\it Mathematical Methods of Classical Mechanics}, 
     Springer-Verlag, New York.}
\refj{Binney, J. \& Tremaine, S.:1987 
     {\it Galactic Dynamics}, 
     Princeton University Press, Princeton NJ. (BT87)}
\refj{Byrd, P.F. \& Friedman, M.D.:1971 
     {\it Handbook of Elliptic Integrals for Engineers and Scientists},
     Springer-Verlag, New York. (BF71)}
\refj{Chandrasekhar, S.:1969 
     {\it Ellipsoidal Figures of Equilibrium}, 
     Dover, New York.}
\refj{Ciotti, L. \& Dutta, S.N.:1994 
     \mnras, {\bf 270}, 390 (CD94)}
\refj{Dehnen, W.:1993 
     \mnras, {\bf 265}, 250}
\refj{Florides, P.S., \& Spyrou, N.K.:1993 
     \apj, {\bf 419}, 541}
\refj{Fong, R., Stevenson, P.R.F., \& Shanks, T.:1990
      \mnras, {\bf 242}, 146}
\refj{Girardi, M., Biviano, A., Giuricin, G., Mardirossian, F., \& 
     Mezzetti, M.:1995
     \apj, {\bf 438}, 527}
\refj{Gradshteyn, I.S. \& Ryzhik, I.M.:1965 
     {\it Tables of Integrals Series and Products}, 
     Academic Press, London. (GR65)}
\refj{Hawley, D.L., \& Peebles, P.J.E.:1975
      \aj, {\bf 80}, 477}
\refj{Kellog, O.D.:1953
     {\it Foundations of Potential Theory},
     Dover, New York.}
\refj{King, I.: 1972 
      \apjl, {\bf 174}, L123}
\refj{MacGillivray, H.T., \& Dodd, R.J.:1979a
      \mnras, {\bf 186}, 69}
\refj{MacGillivray, H.T., \& Dodd, R.J.:1979b
      \mnras, {\bf 186}, 743}
\refj{M\'endez, R.H., Guerrero, M.A., Freeman, K.C., Arnaboldi, M., 
      Kudritzki, P., Hopp, U., \& Capaccioli, M.:1997
      \apjl, {\bf 491}, L23}
\refj{Rhee, G., \& Roos, N.:1990
      \mnras, {\bf 243}, 629}
\refj{Thompson, L.A.:1976
      \apj, {\bf 209}, 22}
\refj{Tremaine, S.D., Richstone, D.O., Byun, Y.I., Dressler, A., Faber, S.M., 
      Grillmair, C., Kormendy, J., \& Lauer, T.R.:1994
      \aj, {\bf 107}, 634}
\refj{Trevese, D., Cirimele, G., \& Flin, P.:1992
      \aj, {\bf 104}, 935}
\refj{Valluri, M.:1993
     \apj, {\bf 408}, 57}
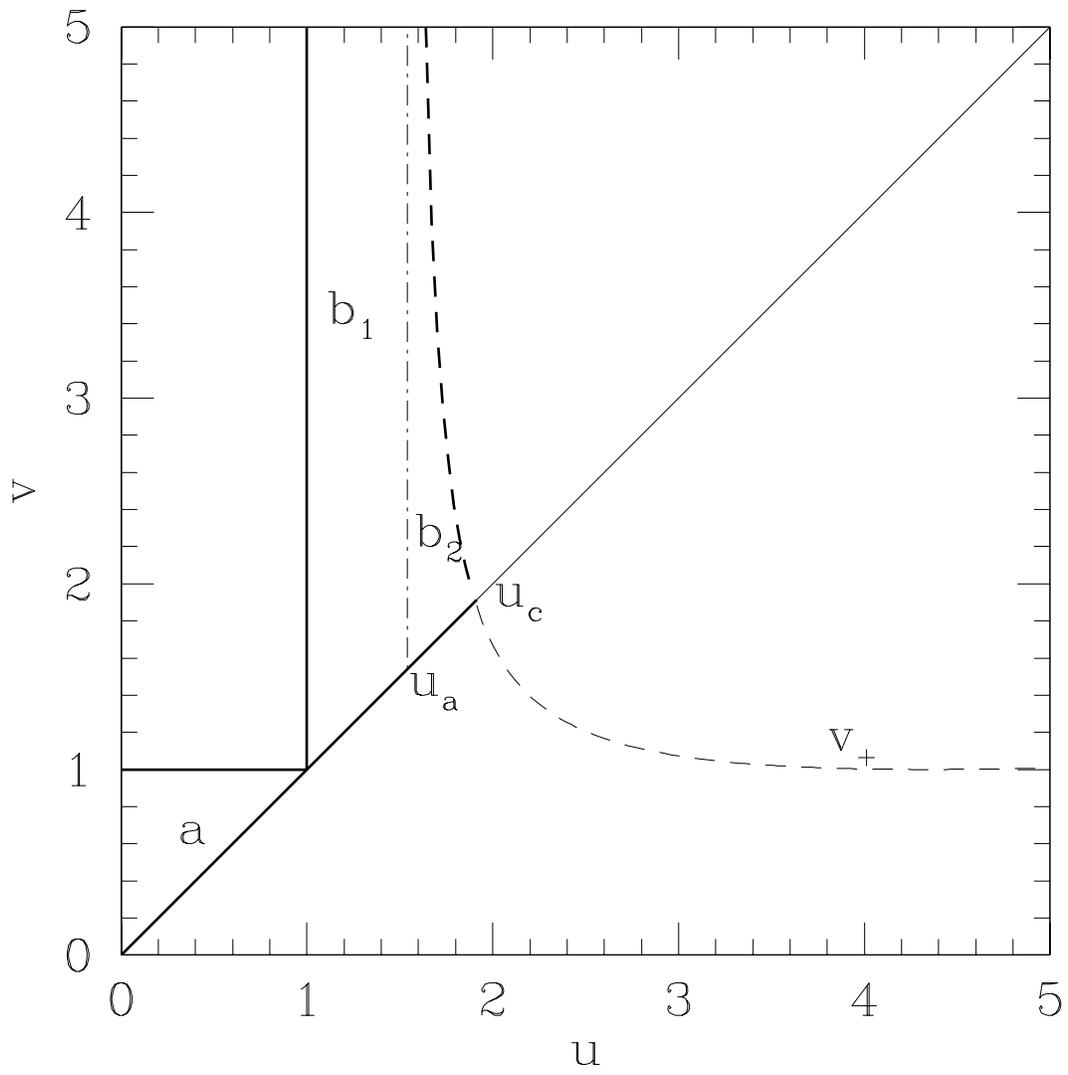
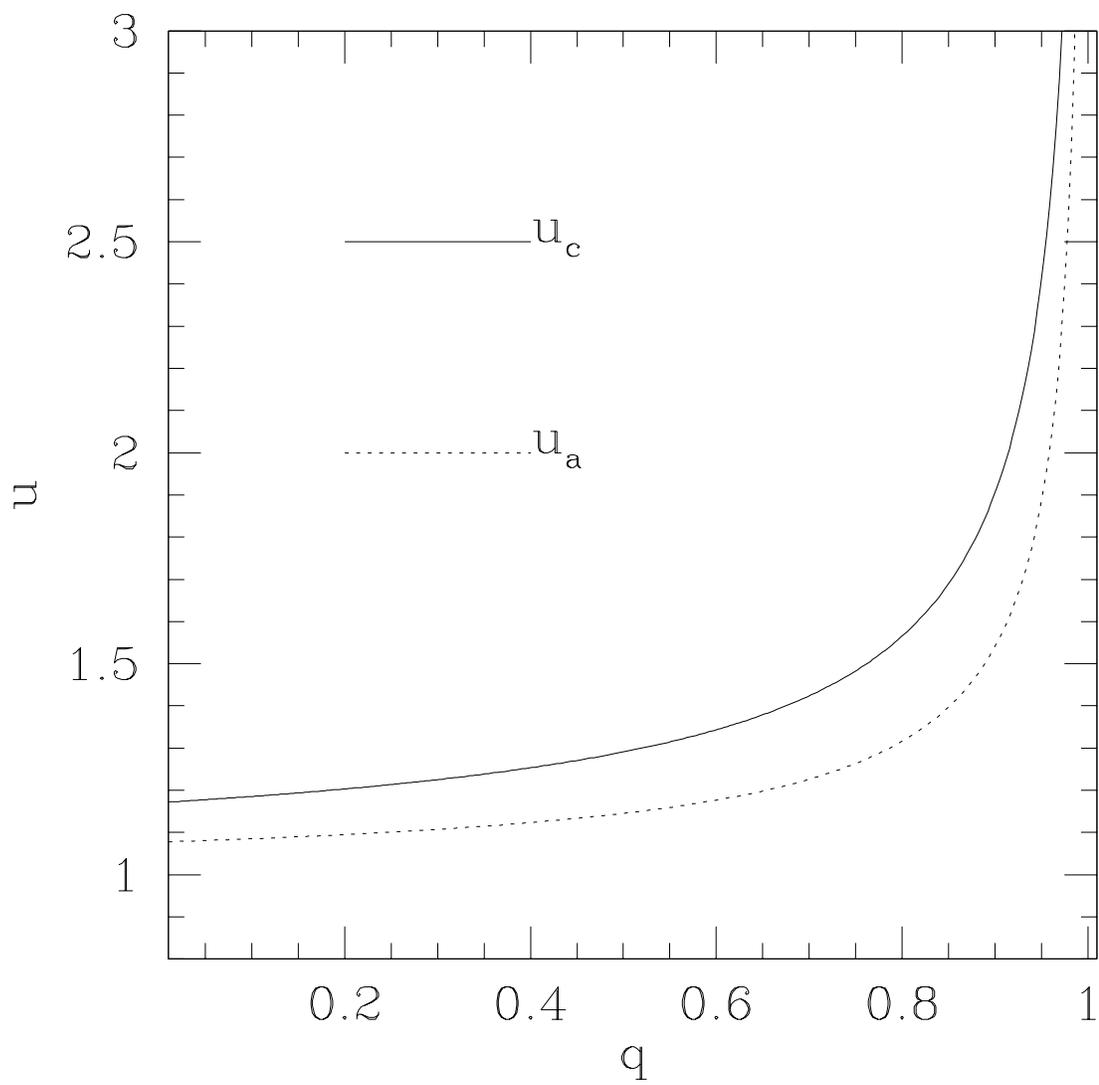
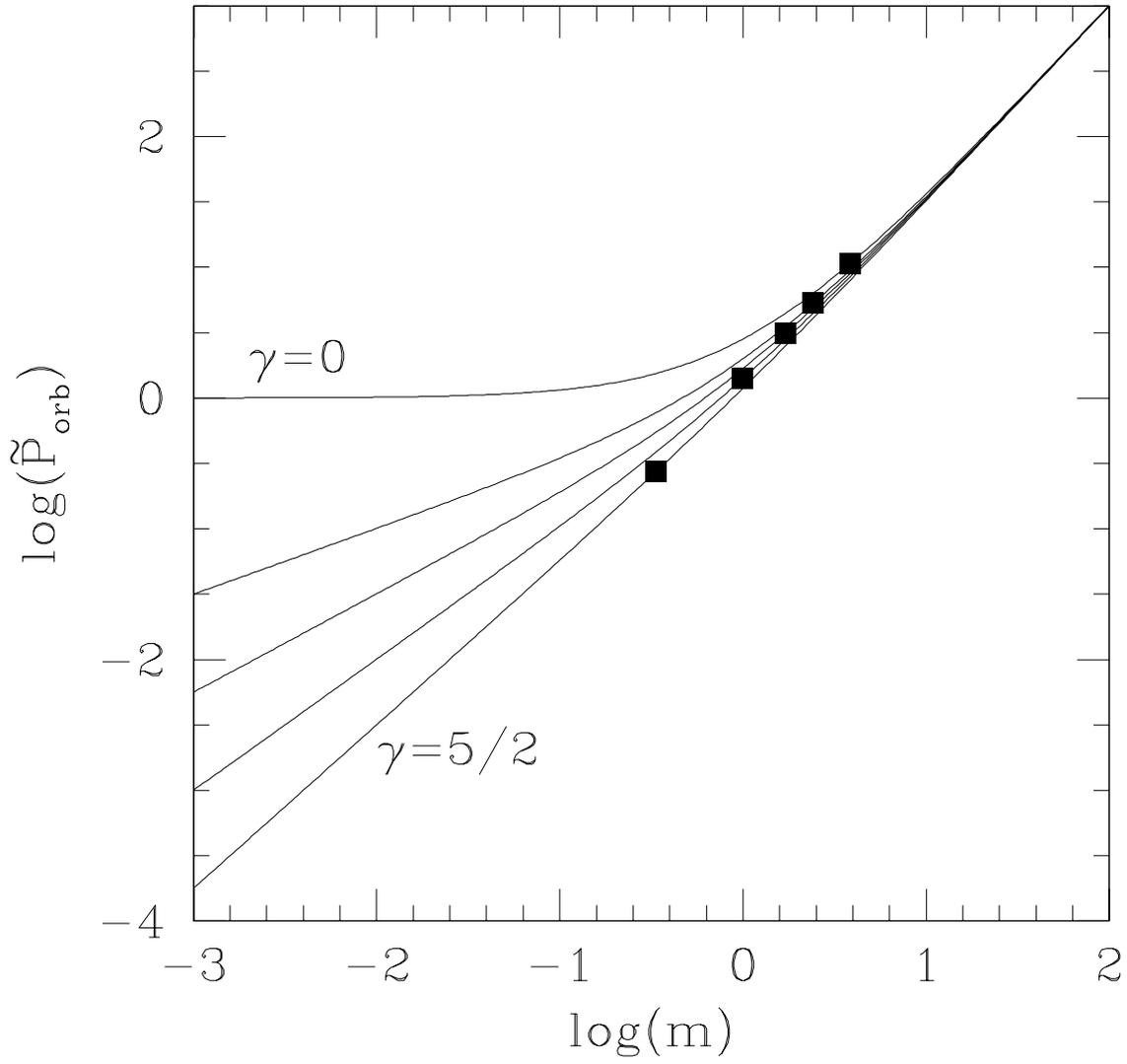
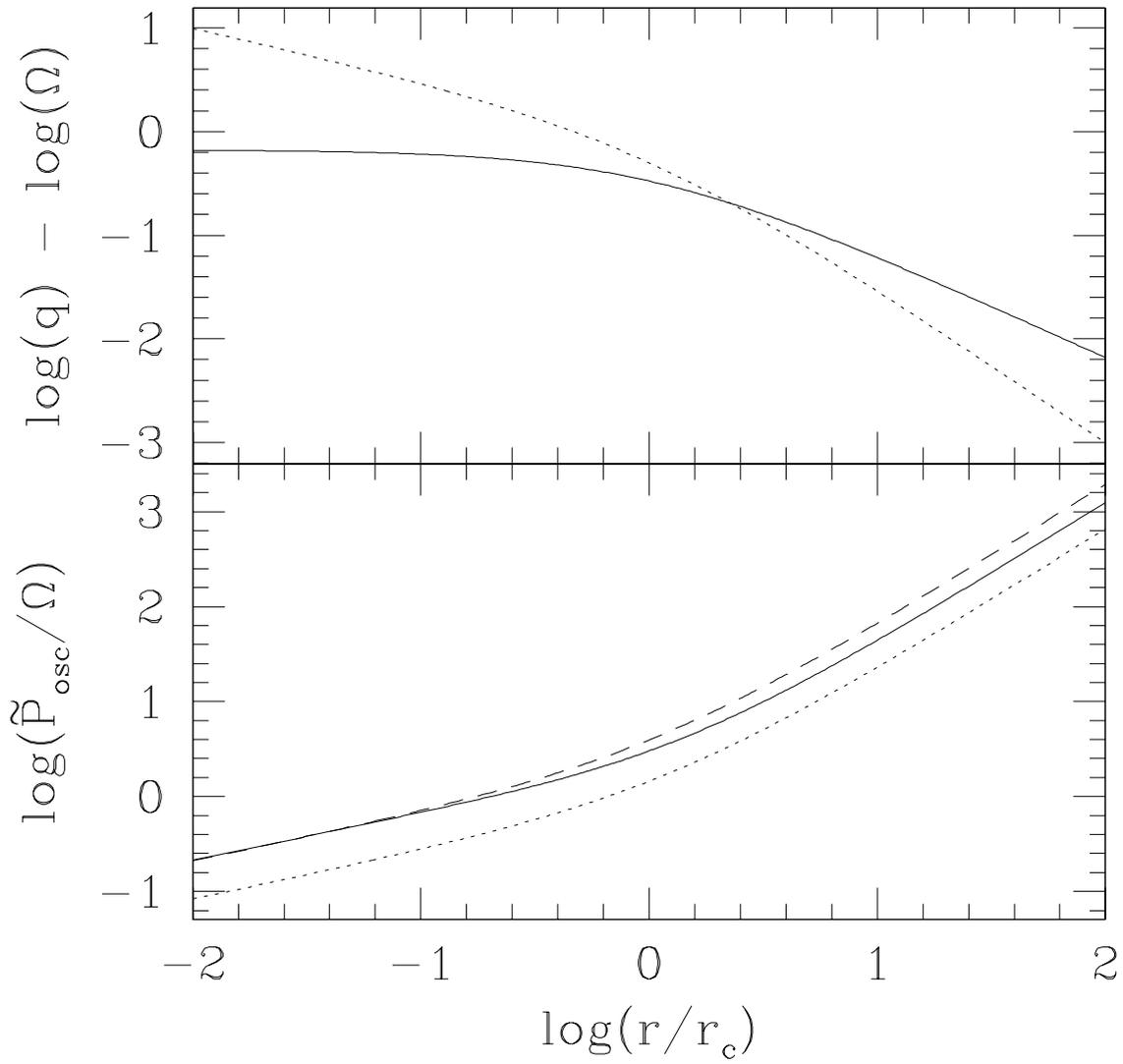
\end{document}